\documentclass[12pt]{article}
\usepackage{a4,epsfig,macros,ukqcd,cite}
\begin{document}

\begin{titlepage}
 
\begin{flushright}
LTH 516
\end{flushright}
 
\vspace*{3mm}
 
\begin{center}
{\Huge Is the up-quark massless?}\\[12mm]
{\large\it UKQCD Collaboration}\\[8mm]

{\bf A.C.~Irving}, {\bf C.~McNeile}, {\bf C.~Michael},
{\bf K.J.~Sharkey} and {\bf H.~Wittig}\footnote{PPARC Advanced Fellow}
\\[3mm] 
Division of Theoretical Physics, Department of Mathematical
Sciences, University of Liverpool, Liverpool L69~3BX, UK\\[3cm]

\end{center}
\vspace{5mm}
\begin{abstract}
We report on determinations of the low-energy constants $\alpha_5$ and
$\alpha_8$ in the effective chiral Lagrangian at O($p^4$), using
lattice simulations with $N_{\rm f}=2$ flavours of dynamical
quarks. Precise knowledge of these constants is required to test the
hypothesis whether or not the up-quark is massless. Our results are
obtained by studying the quark mass dependence of suitably defined
ratios of pseudoscalar meson masses and matrix elements. Although
comparisons with an earlier study in the quenched approximation reveal
small qualitative differences in the quark mass behaviour, numerical
estimates for $\alpha_5$ and $\alpha_8$ show only a weak dependence on
the number of dynamical quark flavours. Our results disfavour the
possibility of a massless up-quark, provided that the quark mass
dependence in the physical three-flavour case is not fundamentally
different from the two-flavour case studied here.
\end{abstract}

\end{titlepage}

\section{Introduction \label{sec_intro}}

A massless up-quark represents a simple and elegant solution to the
strong CP problem. Consequently, the question of whether or not $\mup$
is indeed zero has been the subject of much debate over many years
(for a review see ref.~\cite{BaNiSei94}). Traditionally the problem is
studied in the framework of Chiral Perturbation Theory
(ChPT). Although the most recent estimates point to a non-zero value
for the ratio~$\mup/\mdo$~\cite{leutwyler:1996}, the situation is
complicated by the presence of a hidden symmetry in the effective
chiral Lagrangian~\cite{chir:KapMan86}.  This so-called
``Kaplan-Manohar ambiguity'' implies that~$\mup/\mdo$ can only be
constrained after supplementing ChPT with additional theoretical
assumptions. Although the validity of the commonly used assumptions is
plausible~\cite{chir:Leut90,chir:Leut96}, it is clear that the
question whether the up-quark is massless cannot be studied from first
principles in ChPT.

More recently attention has focussed on lattice simulations to tackle
this problem. A reliable, direct lattice calculation of $\mup$,
however, presents considerable difficulties, even on today's massively
parallel computers. It was therefore proposed to use a more indirect
approach, based on a combination of lattice QCD and
ChPT~\cite{ShaSho_lat99,chir:CoKapNel99,ShaSho_L7}. The aim of this
method is a lattice determination of the so-called ``low-energy
constants'' in the effective chiral Lagrangian, whose precise values
are required to constrain~$\mup/\mdo$ using chiral symmetry and
phenomenological input. A variant of this proposal, which allows for a
determination of the low-energy constants with good statistical
accuracy was discussed in ref.~\cite{mbar:pap4} and tested in the
quenched approximation.

Here we extend the study of~\cite{mbar:pap4} to QCD with two flavours
of dynamical quarks. While this addresses the important issue of
dynamical quark effects, it still does not correspond to the physical
three-flavour case, and thus we are yet unable to give a final answer
to the question in the title of this paper. Nevertheless, our study
represents an important step in an ultimately realistic treatment of
the problem, by studying the dependence of the low-energy constants on
the number of flavours. If our results can be taken over to the
physical case without large modifications -- and there are indications
that this is not unreasonable -- then the possibility of a massless
up-quark is strongly disfavoured.

In Section~\ref{sec_lec} we briefly review the Kaplan-Manohar
ambiguity and its relevance for the value of
$\mup$. Section~\ref{sec_lat} contains details of our lattice
simulations, whose results are described in Section~\ref{sec_res}. In
Section~\ref{sec_disc} we discuss the implications of our findings and
present an outlook to future work.

\section{Low-energy constants and $\mup=0$ \label{sec_lec}}

In order to make this paper self-contained, we briefly review the
implications of the Kaplan-Manohar ambiguity for the ratio
$\mup/\mdo$. The strategy to address the problem in lattice
simulations will then become clear. A more complete discussion can be
found in
refs.~\cite{leutwyler:1996,chir:Leut96,chir:CoKapNel99,mbar:pap4}.

A determination of $\mup/\mdo$ in ChPT which is able to distinguish
between a massless and a massive up-quark requires precise knowledge
of the first-order mass correction term $\Delta_{\rm M}$. At
order~$p^4$ in the chiral Lagrangian it is given
by~\cite{chir:GaLe1,chir:GaLe2,chir:param}
\be
    \Delta_{\rm M} = \frac{\mK^2-\mpi^2}{(4\pi\Fpi)^2}\,
    (2\alpha_8-\alpha_5) + \hbox{chiral logs},
\ee
where $\Fpi=93.3\,\mev$ is the pion decay constant, and $\alpha_5$,
$\alpha_8$ are low-energy constants, whose values have to be
determined from phenomenology. Throughout this paper we adopt a
convention in which the low-energy constants $\alpha_i$ are related to
the corresponding constants $L_i$ of ref.~\cite{chir:GaLe2} through
$\alpha_i=8(4\pi)^2L_i$. Furthermore, we always quote low-energy
constants in the $\msbar$-scheme at scale $\mu=4\pi\Fpi$.

The value of $\alpha_5$ can be extracted from the ratio of
pseudoscalar decay constants, $\FK/\Fpi$, and is given by
\be
    \alpha_5 = 0.5\pm0.6.
\label{eq_a5_phen}
\ee
By contrast, there is no direct phenomenological information on
$\alpha_8$ or the combination $(2\alpha_8-\alpha_5)$. Although
$\alpha_8$ is contained in the correction to the Gell-Mann--Okubo
formula, i.e.
\be
   \Delta_{\rm GMO} = \frac{\mK^2-\mpi^2}{(4\pi\Fpi)^2}\,
   (\alpha_5-12\alpha_7-6\alpha_8) + \hbox{chiral logs},
\ee
its determination requires prior knowledge of $\alpha_7$. It is at
this point that the Kaplan-Manohar (KM) ambiguity becomes
important. It arises from the observation that a simultaneous
transformation of the quark masses
\be
  \mup\to\mup+\lambda\mdo\mst,\quad
  \mdo\to\mdo+\lambda\mst\mup,\quad
  \mst\to\mst+\lambda\mup\mdo,
\label{eq_KM_mass}
\ee
and coupling constants according to
\be
  \alpha_6\to\alpha_6+\lambda\frac{(4\pi F_0)^2}{4B_0},\quad
  \alpha_7\to\alpha_7+\lambda\frac{(4\pi F_0)^2}{4B_0},\quad
  \alpha_8\to\alpha_8-\lambda\frac{(4\pi F_0)^2}{2B_0},
\label{eq_KM_coup}
\ee
leaves the effective chiral Lagrangian invariant. Here, $\lambda$ is
an arbitrary parameter, and $F_0$, $B_0$ are coupling constants in the
lowest-order chiral Lagrangian.\footnote{$F_0$ coincides with $\Fpi$
at lowest order.} Thus, chiral symmetry cannot distinguish between
different sets of quark masses and coupling constants, which are
related through eqs.~(\ref{eq_KM_mass}) and~(\ref{eq_KM_coup}). Indeed
the correction $ \Delta_{\rm GMO}$ is invariant under the above
transformations. The value of $\alpha_7$ can be fixed by invoking
additional theoretical assumptions, such as the validity of
large-$N_c$ arguments for $N_c=3$. In accordance with these
assumptions, Leutwyler~\cite{leutwyler:1996} constrained the
correction term $\Delta_{\rm M}$ to be small and positive:
\be
   0<\Delta_{\rm M}\leq0.13.
\label{eq_DeltaM_est}
\ee
This gives $\mup/\mdo=0.553(43)$ and hence a non-zero value for
$\mup$. The ``standard'' values for $\alpha_7$ and $\alpha_8$ which
are compatible with \eq{eq_DeltaM_est}
are~\cite{chir:GaLe1,chir:GaLe2,chir:param}
\be
   \alpha_7 = -0.5 \pm0.25,\qquad \alpha_8 =  0.76\pm0.4.
\label{eq_a7a8_std}
\ee
In view of the importance of the strong CP problem, one may regard any
analysis based on theoretical assumptions beyond chiral symmetry as
insufficient. In particular, since the uncertainties in the estimates
for~$\alpha_7$ and~$\alpha_8$ are quite large, the possibility that
$\mup=0$ does not appear to be ruled out completely. A massless
up-quark would require~\cite{chir:CoKapNel99}
\be
   \alpha_7 = 0.25 \pm0.25,\qquad \alpha_8 =  -0.9\pm0.4,
\label{eq_a7a8_mup0}
\ee
resulting in a large, negative first-order correction
$\Delta_{\rm{M}}$. In order to decide which scenario is realised and
to pin down the value of $\mup$ one has to replace the theoretical
assumptions by a solution of the underlying theory of QCD.

The KM ambiguity implies that the low-energy constants $ \alpha_7$,
$\alpha_8$ (and $ \alpha_6$) can be determined from chiral symmetry
and phenomenology only if the physical quark masses are known
already. At this point it is important to realise that QCD is not
afflicted with the KM ambiguity, and that the formalism of ChPT also
holds for unphysical quark masses. Since quark masses are input
parameters in lattice simulations of QCD, their relations to hadronic
observables need not be known {\it a priori}. Hence, the low-energy
constants can be determined by studying pseudoscalar meson masses and
matrix elements for unphysical quark masses and fitting their quark
mass dependence to the expressions found in ChPT. In this way it is
possible to determine $\alpha_5$ and -- more importantly -- the
combination $(2\alpha_8-\alpha_5)$ directly in lattice simulations.

\section{Lattice setup and simulation details\label{sec_lat}}

In ref.~\cite{mbar:pap4} it was shown how the low-energy constants can
be extracted from lattice data for suitably defined ratios of
pseudoscalar masses and matrix elements, $\RFG$ and $\RF$. Here we
repeat their definitions in order to explain the necessary
notation. For more details we refer the reader to the original
paper~\cite{mbar:pap4}.

The actual determination of the low-energy constants proceeds by
studying the mass dependence of $\RFG$ and $\RF$ around some reference
quark mass $\mref$. As pointed out before, $\mref$ does not have to
coincide with a physical quark mass~\cite{mbar:pap3}, as long as it is
small enough for ChPT to be applicable.

In this paper we will be concerned with simulations of ``partially
quenched'' QCD, where valence and sea quarks can have different
masses. Let us therefore consider a pseudoscalar meson with valence
quark masses $m_1$ and $m_2$ at a fixed value of the sea quark mass
$\msea$. In \cite{mbar:pap4} the dimensionless mass parameters
\be
   y=\frac{B_0(m_1+m_2)}{(4\pi F_0)^2},\quad
  \yref=\frac{2B_0\mref}{(4\pi F_0)^2}, \quad x=y/\yref,
\ee
were defined, as well as the ratios
\bea
  \RFG(x) &=& \left(\frac{\Fp(y)}{\Gp(y)}\right)\left/
  \left(\frac{\Fp(\yref)}{\Gp(\yref)}\right)
  \right. \nonumber \\
  \RF(x)  &=& \Fp(y)\left/\Fp(\yref)\right.,\qquad x=y/\yref.
\label{eq_RFG_def1} 
\eea
Here $\Fp$ is the pseudoscalar decay constant, and $\Gp$ is the matrix
element of the pseudoscalar density between a pseudoscalar state and
the vacuum. The parameter $x$ denotes the fraction of the reference
quark mass at which the ratios $\RFG$ and $\RF$ are considered.

As emphasised in~\cite{mbar:pap4}, $\RFG$ and $\RF$ are well-suited to
extract the low-energy constants, since ratios are usually obtained
with high statistical accuracy in numerical simulations. Furthermore,
any renormalisation factors associated with $\Fp$ and $\Gp$ drop out,
so that $\RFG$ and $\RF$ can be readily extrapolated to the continuum
for every fixed value of~$x$. As has been shown in the quenched
approximation~\cite{mbar:pap4}, discretisation errors in $\RFG$ and
$\RF$ are very small, so that good control over lattice artefacts is
achieved. This issue is important, since the determination of
low-energy constants requires an unambiguous separation of the mass
dependence from effects of non-zero lattice spacing.

We now give the expressions for $\RFG$ and $\RF$ in partially quenched
QCD which are relevant for our study. They were obtained using the
results of ref.~\cite{chir:Sha97} and are listed in appendix~A
of~\cite{mbar:pap4}. Our reference point $\yref$ was always defined at
\be
  m_1=m_2=\msea=\mref,\quad \yref=\frac{2B_0\mref}{(4\pi F_0)^2}.
\label{eq_yref_PQ}
\ee
In order to map out the mass dependence of $\RFG$ and $\RF$ for a
fixed value of $\msea$, we have considered the cases labelled ``VV''
and ``VS1'' in ref.~\cite{mbar:pap4}. The first uses degenerate
valence quarks and is defined by
\be
   \hbox{VV:}\quad m_1=m_2=x\mref,\quad \msea=\mref,
\label{eq_xdep_VV}
\ee
which leads to the expressions
\bea
\RFG^{\rm VV}(x) &=&
   1-\frac{1}{\Nf}\yref\left[(2x-1)\ln{x}+2(x-1)\ln\yref\right]
   \nonumber\\
   & &
   -\yref(x-1)\left[(2\alpha_8-\alpha_5)+\textstyle\frac{1}{\Nf}\right]
  \label{eq_RFG_VV}  \\
\RF^{\rm VV}(x) &=&
   1-\frac{\Nf}{4}\yref\left[
   (x+1)\ln\left(\textstyle\frac{1}{2}(x+1)\right)
   +(x-1)\ln\yref\right]
   \nonumber\\
   & &+\yref(x-1)\textstyle\frac{1}{2}\alpha_5,
   \label{eq_RFPS_VV}
\eea
where $\Nf$ denotes the number of dynamical quark flavours. The case
labelled ``VS1'', based on non-degenerate valence quarks, is defined
by
\be
   \hbox{VS1:}\quad m_1=x\mref,\quad m_2=\msea=\mref.
\label{eq_xdep_VS1}
\ee
According to Table~1 of ref.~\cite{mbar:pap4} the expressions for the
ratios are then given by
\bea
\RFG^{\rm VS1}(x) &=&
   1-\frac{1}{\Nf}\yref\left[x\ln{x}+(x-1)\ln\yref\right]
   -\yref(x-1)\textstyle\frac{1}{2}(2\alpha_8-\alpha_5)
  \label{eq_RFG_VS1}  \\
\RF^{\rm VS1}(x) &=&
   1-\frac{\Nf}{8}\yref\left[
   (x+1)\ln\left(\textstyle\frac{1}{2}(x+1)\right)
   +(x-1)\ln\yref +\textstyle\frac{2}{\Nf^2}\ln{x}\right] \nonumber\\
   & &+\yref(x-1)\textstyle\frac{1}{4}
      \left(\alpha_5+\textstyle\frac{1}{\Nf}\right). 
   \label{eq_RFPS_VS1}
\eea

Our simulations were performed for $\Nf=2$ flavours of dynamical,
O($a$) improved Wilson fermions. The value of the bare coupling was
set to $\beta=6/g_0^2=5.2$. The improvement coefficient $\csw$, which
multiplies the Sheikholeslami-Wohlert term in the fermionic action,
was taken from the interpolating formula of
ref.~\cite{impr:csw_nf2}. Here we considered a single value of the sea
quark mass, corresponding to a hopping parameter $\ksea=0.1355$. For
this choice of parameters we generated 208 dynamical gauge
configurations on a lattice of size $16^3\cdot32$. For further details
we refer to
\cite{dspect:ukqcd_lat99,dspect:ukqcd_lat00,dspect:ukqcd_csw202}. Here
we only mention that the hadronic radius $r_0$ defined through the
force between static sources~\cite{pot:r0} has been determined as
$r_0/a=5.041(40)$~\cite{dspect:ukqcd_csw202}. For $r_0=0.5\,\fm$ this
implies that the lattice spacing in physical units is
$a=0.099(1)\,\fm$.

We have computed quark propagators for valence quarks with hopping
parameters $\kval=0.1340$, 0.1345, 0.1350, 0.1355 and 0.1358. In
addition to local operators, we have also calculated propagators for
fuzzed sinks and/or sources, using the procedure
of~\cite{ukqcd:fuzz}. In the pseudoscalar channel we employed both the
pseudoscalar density and the temporal component of the axial current
as interpolating operators. These two types were used together with
the different combinations of fuzzed and local propagators to
construct a $4\times4$ matrix correlator for pseudoscalar mesons. By
performing factorising fits using an ansatz that incorporates the
ground state and the first excitation, we were able to extract the
pseudoscalar mass $\mps$, as well as the matrix elements of the axial
current and the pseudoscalar density, i.e.
\be
   \zeta_{\rm A}=\langle0|A_0|{\hbox{PS}}\rangle,\qquad
   \zeta_{\rm P}=\langle0| P |{\hbox{PS}}\rangle.
\ee
In order to be consistent with O($a$) improvement the amplitudes
$\zeta_{\rm A}$ and $\zeta_{\rm P}$ must be related to the matrix
elements of the improved currents and densities at non-zero quark
mass. Using the definitions of~\cite{impr:pap1} it is then easy to see
that the (unrenormalised) pseudoscalar decay constant is given by
\be
   \Fp = (1+\ba{a\mq})\left\{\frac{\zeta_{\rm A}}{\mps}
        +\ca\frac{\zeta_{\rm P}}{\mps}\sinh(a\mps)\right\},
\label{eq_Fp_def}
\ee
whereas $\Gp$ is given by
\be
   \Gp = (1+\bp{a\mq})\zeta_{\rm P}.
\label{eq_Gp_def}
\ee
The improvement coefficients $\ba$, $\bp$ and $\ca$ were computed in
one-loop perturbation theory~\cite{impr:pap2,impr:pap5} in the bare
coupling, since non-perturbative estimates are not available at
present. For the improvement coefficient $\ca$ this procedure yields
\be
    \ca=-0.0087.
\ee
For degenerate valence quarks the quark mass $\mq$ is given by
\be
   a\mq = \frac{1}{2}\left(\frac{1}{\kval}-\frac{1}{\kcrit}\right),
\ee
where we have inserted $\kcrit=0.13693$ for the critical hopping
parameter, as estimated in \cite{impr:csw_nf2}. Finally we note that
the current quark mass~$m$ defined through the PCAC relation in the
O($a$) improved theory is obtained from $\zeta_{\rm A}$, $\zeta_{\rm
P}$ and~$\mps$ via
\be
   am = \frac{\zeta_{\rm A}}{2\zeta_{\rm P}}\sinh(a\mps)
       +\frac{1}{2}\ca\sinh^2(a\mps).
\label{eq_mPCAC_def}
\ee
All our statistical errors were obtained using a bootstrap
procedure~\cite{efron}.

\section{Results \label{sec_res}}

Our results for pseudoscalar masses, matrix elements and current quark
masses are listed in Table~\ref{tab_rawdata}. Compared
with~\cite{dspect:ukqcd_csw202} the numbers for the pseudoscalar
masses reported here may differ by up to one standard deviation, as a
result of using a larger matrix correlator in the fitting procedure.

Following \eq{eq_yref_PQ} we have defined the mass $\mref$ at the
reference point for $\kappa_1^{\rm val}=\kappa_2^{\rm{val}}=\ksea =
0.1355$, which corresponds to $\left.(r_0\mps)^2\right|_{m=\mref}
=2.092$. Using the leading-order relations $\mps^2=2{B_0}m$ and
$F_0=\Fpi=93.3\,\mev$ this implies
\be
  \yref=0.2370.
\ee
It is instructive to compare these values with those of the previous
quenched study~\cite{mbar:pap4}, where $\yref=0.3398$,
$(r_0\mps)^2|_{m=\mref}=3$, and $\mref\approx\mst$. Hence, the value
of $\mref$ employed in this paper is smaller by 30\%, such that
$\mref\approx0.7\mst$. A smaller value of $\mref$ is clearly desirable
for our purpose, since the predictions of ChPT are expected to hold
more firmly for smaller masses. Nevertheless, our sea quarks are still
relatively heavy, as signified by the ratio $(\mps/m_{\rm
V})|_{m=\msea}\approx0.58$~\cite{dspect:ukqcd_csw202}, which is to be
compared to $m_\pi/m_\rho=0.169$.

In the spirit of partially quenched QCD we have considered valence
quarks that are lighter than the sea quarks. We were thus able to
extend the quark mass range down to about $\mst/2$, which is quite a
bit below the smallest mass reached in \cite{mbar:pap4}. This may be
an indication that the inclusion of dynamical quarks alleviates the
problem of exceptional configurations, which precludes attempts to
work at very small quark masses in the quenched
approximation. However, our attempts to push to valence quark masses
below $\mst/2$ have proved unsuccessful, due to the appearance of
exceptional configurations for $\mval/\msea\;\lesssim\;0.7$. More
precisely, we observed large statistical fluctuations in hadron
correlators computed from quark propagators at
$\kappa^{\rm{val}}=0.1360$, despite the fact that the inversion
algorithm converged.

\begin{table}
\begin{center}
\begin{tabular}{c ccc c c c}
\hline
\hline \\[-1.0ex]
$\kappa_1^{\rm val}$ & $\kappa^{\rm val}_2$ && $a\mps$
  & $a\zeta_{\rm A}/\mps$ & $a^2\zeta_{\rm P}$ &  $am$  \\[1.0ex]
\hline \\[-1.0ex]
0.1358 & 0.1358 && 0.2301\err{52}{51} & 0.0995\err{24}{30} 
 & 0.1782\err{70}{57} & 0.0147\er{4}{5} \\
0.1355 & 0.1355 && 0.2869\err{40}{41} & 0.1068\err{22}{25}
 & 0.1893\err{57}{52} & 0.0232\er{4}{5} \\
0.1350 & 0.1350 && 0.3585\err{26}{28} & 0.1166\err{20}{24}
 & 0.1046\err{47}{49} & 0.0368\er{4}{5} \\
0.1345 & 0.1345 && 0.4192\err{20}{23} & 0.1246\err{19}{24}
 & 0.2187\err{40}{45} & 0.0507\er{5}{5} \\
0.1340 & 0.1340 && 0.4739\err{17}{20} & 0.1312\err{17}{25}
 & 0.2315\err{43}{44} & 0.0650\er{6}{5} \\
\hline
0.1358 & 0.1355 && 0.2607\err{45}{46} & 0.1033\err{24}{26}
 & 0.1841\err{60}{55} & 0.0190\er{4}{4} \\
0.1350 & 0.1355 && 0.3249\err{30}{37} & 0.1118\err{20}{25}
 & 0.1970\err{50}{52} & 0.0300\er{4}{5} \\
0.1345 & 0.1355 && 0.3591\err{27}{32} & 0.1158\err{20}{24}
 & 0.2036\err{50}{46} & 0.0369\er{4}{5} \\
0.1340 & 0.1355 && 0.3907\err{25}{28} & 0.1190\err{21}{25}
 & 0.2094\err{44}{45} & 0.0438\er{4}{6} \\[1.0ex]
\hline
\hline
\end{tabular}
\end{center}
\caption{\small Results for pseudoscalar masses, matrix elements and
current quark masses at $\beta=5.2$,
$\ksea=0.1355$.\label{tab_rawdata}}
\end{table}

The results in Table~\ref{tab_rawdata} can now be used to compute
$\RFG$ and $\RF$ through eqs.~(\ref{eq_Gp_def}), (\ref{eq_Fp_def})
and~(\ref{eq_RFG_def1}). In order to map out their quark mass
dependence in some detail, we have performed local interpolations of
the results for $\Fp$ and $\Fp/\Gp$ to 20 different values of the
dimensionless mass parameter~$x$ in the range $0.7\leq{x}\leq2.6$,
separated by increments of~0.1.

Since we only have one $\beta$-value we cannot extrapolate $\RFG$,
$\RF$ to the continuum limit for fixed~$x$. Unlike
ref.~\cite{mbar:pap4}, where such extrapolations could be performed,
we must compare our data to ChPT at non-zero lattice spacing. Our
estimates for the low-energy constants are therefore subject to an
unknown discretisation error. It is reasonable to assume, however,
that cutoff effects in $\RFG$ and $\RF$ are fairly small, owing to
cancellations of lattice artefacts of similar size between numerator
and denominator. We will return to this issue below, when we discuss
our final estimates for the low-energy constants.

\begin{figure}[tb]
\hspace{0cm}
\vspace{-3.cm}

\centerline{
\leavevmode
\psfig{file=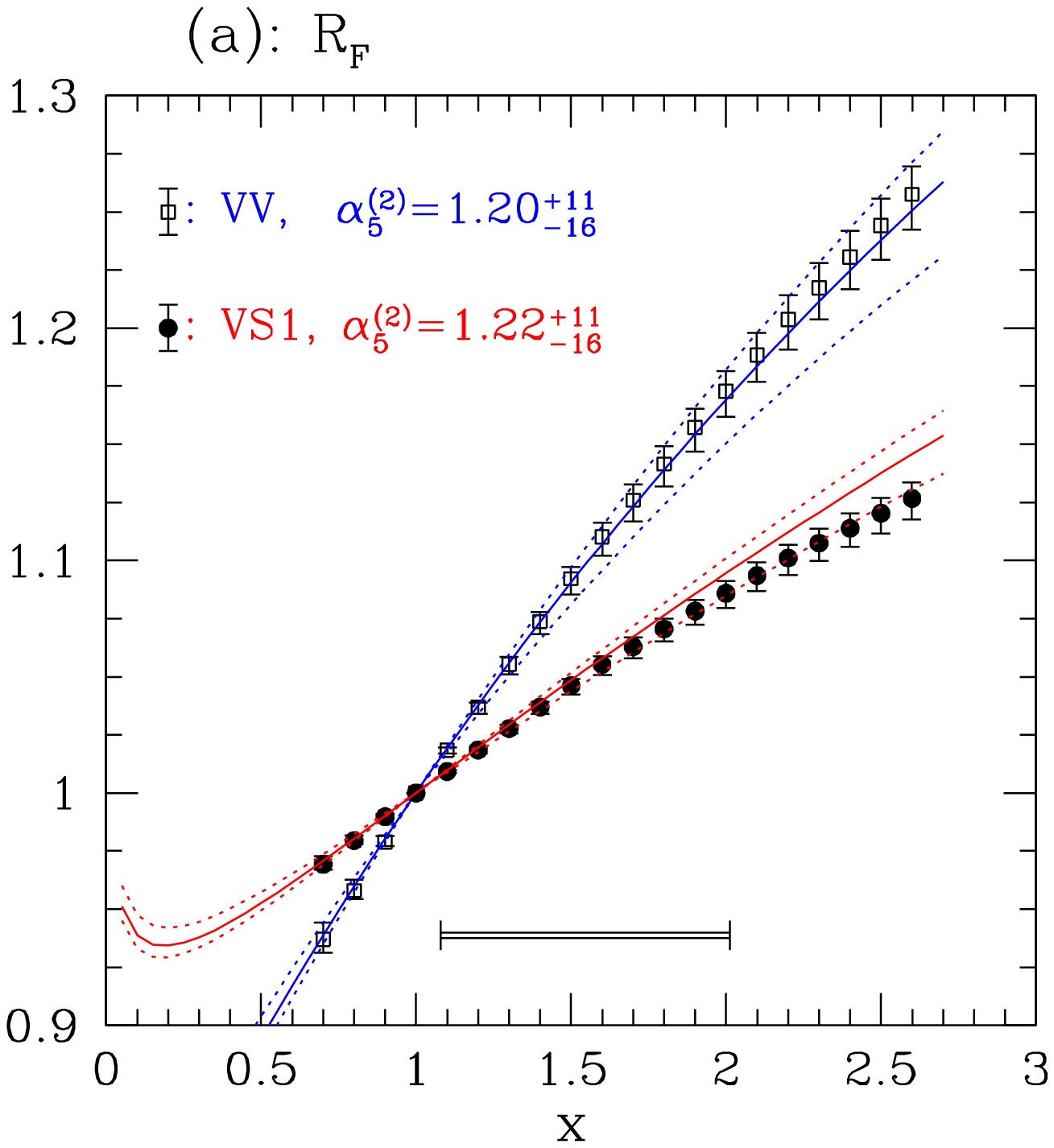,width=12cm}
\leavevmode
\hspace{-4.0cm}
\psfig{file=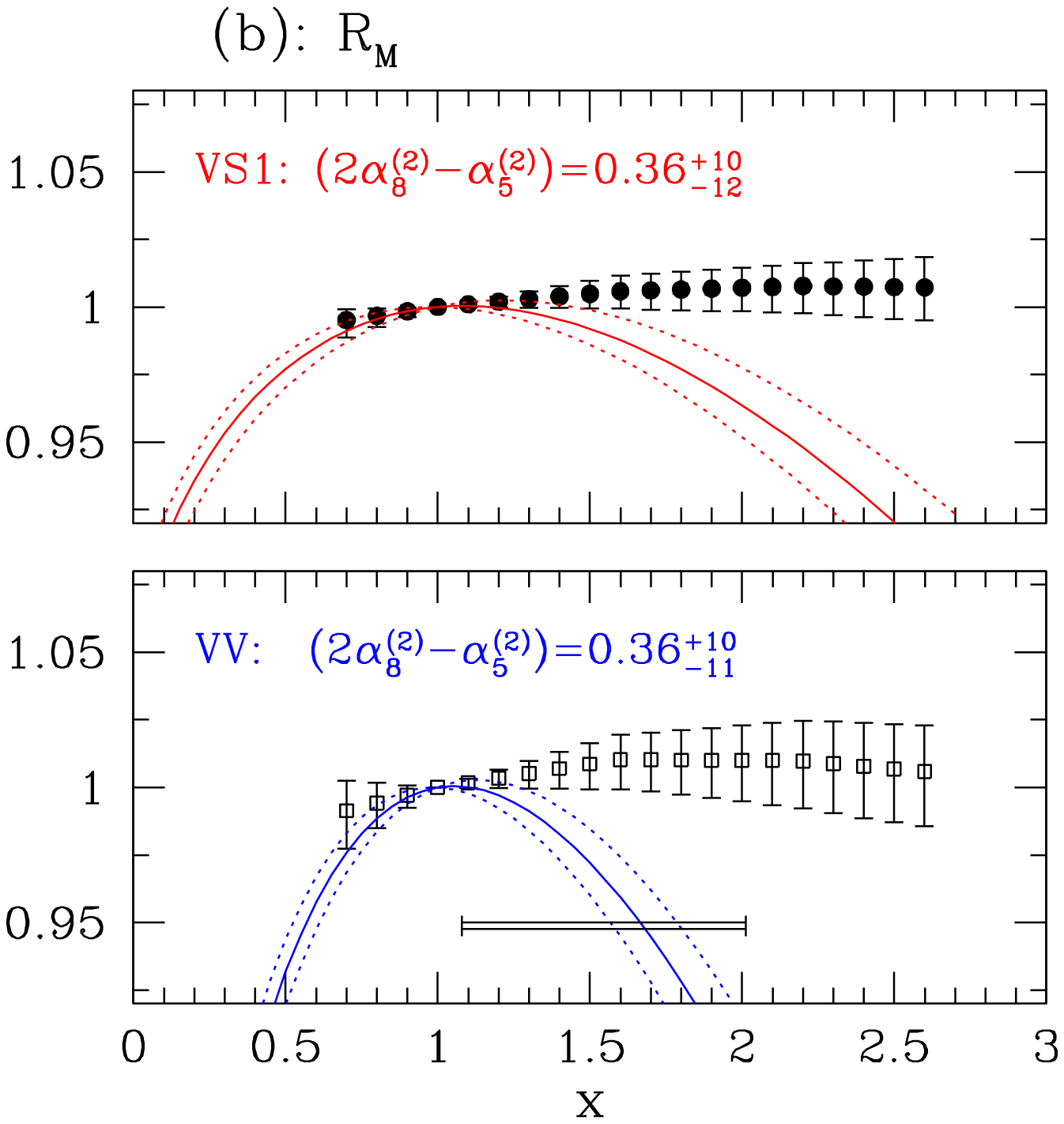,width=12cm}
}
\vspace{-0.5cm}
\renewcommand{\baselinestretch}{0.95}
\caption{\small (a): The ratios $\RF$ for the cases labelled ``VV''
and ``VS1'' compared with the fit to the expressions in ChPT; (b): the
same for the ratios $\RFG$. Dotted lines indicate the variation due to
the statistical uncertainty in the low-energy constants. The straight
double lines indicate the quark mass range explored in the earlier
quenched study~\protect\cite{mbar:pap4}.
\label{fig_results}}
\end{figure}

In order to extract the low-energy constants we have restricted the
$x$-interval to $0.7\leq{x}\leq1.1$, thereby seeking to maximise the
overlap with the domain of applicability of ChPT, whilst maintaining a
large enough interval to check the stability of our results. Estimates
for the low-energy constants were obtained by fitting the data for
$\RFG$ and $\RF$ to the corresponding expressions for the ``VV'' and
``VS1'' cases listed in Section~\ref{sec_lat}. Since $\RF$ is linear
in $\alpha_5$ one can obtain this low-energy constant also from simple
algebraic expressions involving the difference $\RF(x_1)-\RF(x_2)$ for
two distinct arguments, $x_1$ and~$x_2$. A similar relation can be
used to compute $(2\alpha_8-\alpha_5)$ from $\RFG(x_1)-\RFG(x_2)$. We
have checked that both methods give consistent results and obtain
\bea
   \hbox{VV:}  && \alpha_5^{(2)} = 1.20\err{11}{16},\quad
   (2\alpha_8^{(2)}-\alpha_5^{(2)}) = 0.36\err{10}{10}, \\
   \hbox{VS1:} && \alpha_5^{(2)} = 1.22\err{11}{16},\quad
   (2\alpha_8^{(2)}-\alpha_5^{(2)}) = 0.36\err{10}{12},
\eea
where the errors are purely statistical. From here on we also indicate
the number of dynamical quark flavours as a superscript, to
distinguish these estimates from the corresponding ones in the
quenched and three-flavour cases. Our results can now be inserted back
into the expressions for $\RFG$ and $\RF$. The resulting curves are
plotted together with the data in Fig.~\ref{fig_results}.

It is striking that the data for $\RF$ in the VV case are described
remarkably well over the whole mass range, despite the fact that
$\alpha_5^{(2)}$ has only been determined for $x\leq1.1$.  The
qualitative behaviour of $\RF$ -- which features a slight curvature --
is thus rather well modelled by eqs.~(\ref{eq_RFPS_VV})
and~(\ref{eq_RFPS_VS1}), which include a linear term as well as chiral
logarithms. By contrast, there are no logarithmic contributions to
$\RF$ in quenched ChPT, and the expected purely linear behaviour has
indeed been observed in the data~\cite{mbar:pap4}. Of course,
higher-order terms in the quark mass could in principle produce a
curvature, and therefore these observations do not provide unambiguous
evidence for chiral logarithms. Nevertheless, it is remarkable that
the clear distinction between the expressions for $\RF$ in partially
quenched and quenched ChPT (i.e. the presence, respectively absence of
chiral logarithms) is accompanied by corresponding qualitative
differences in the numerical data.

Unlike $\RF$ the ratio $\RFG$ is not described well for larger masses,
which may signal a breakdown of the chiral expansion for this quantity
for masses not much larger than $\mst$. It is therefore conceivable
that higher orders in ChPT affect the extraction of
$(2\alpha_8^{(2)}-\alpha_5^{(2)})$. However, without access to quark
masses that are substantially lower than our simulated ones, it is not
easy to quantify reliably any uncertainty due to neglecting higher
orders. 

One way to examine the influence of higher orders is to extract the
low-energy constants from a mass interval of fixed length,
$(x_{\rm{max}}-x_{\rm{min}})=0.4$, which is then shifted inside an
extended range of $0.7\leq{x}\leq1.5$. The spread of results so
obtained then serves as an estimate of the systematic errors incurred
by neglecting higher orders. We note that $x=1.5$ corresponds to a
quark mass slightly larger than $\mst$. For $\alpha_5^{(2)}$ such a
procedure yields only a small variation of $\pm0.05$. This is not
surprising, since $\RF$ is modelled very well over the entire mass
range. By contrast, the spread of results obtained for
$(2\alpha_8^{(2)}-\alpha_5^{(2)})$ is as large as $\pm0.15$. Whether
or not these numbers represent realistic estimates of the actual
uncertainty cannot be decided at this stage. In order to be more
conservative we have decided to quote a systematic error of $\pm0.2$
for {\it all\/} low-energy constants. We note that this level of
uncertainty due to neglecting higher orders was also quoted in the
quenched case~\cite{mbar:pap4}, where quark masses were slightly
larger.

Since we do not have enough data to extrapolate $\RF$ and $\RFG$ to
the continuum limit we also have to estimate a systematic error due to
cutoff effects. As explained above, however, we expect such effects to
be small. In order to get an idea of the typical size of
discretisation errors we have looked again at quenched data obtained
at $\beta=5.93$
\cite{dspect:ukqcd_lat99,dspect:ukqcd_lat00,dspect:ukqcd_csw202}
and~6.0~\cite{mbar:pap4}, for which the lattice spacing in physical
units is roughly the same as in our dynamical simulations
($a\approx0.1\,\fm$). For both $\beta=5.93$ and~6.0 the results for
$\RFG$ and $\RF$ are mostly consistent within errors with the
corresponding values in the continuum limit (see, for instance, Fig.~1
in~\cite{mbar:pap4}). Furthermore, low-energy constants extracted for
$a\approx0.1\,\fm$ differ from the results in the continuum limit by
less than one standard deviation. Although these findings cannot be
taken over literally to the dynamical case without direct
verification, they nevertheless indicate that lattice artefacts are
small enough such that does not have to expect large distortions in
our estimates for the low-energy constants. In order to take account
of these observations we have decided to quote an additional
systematic error due to lattice artefacts, which is as large as the
statistical error.

Since non-perturbative estimates for the improvement coefficients
$\ba$, $\bp$ and~$\ca$ are not available for $\Nf=2$, one may be
worried that there are large uncancelled lattice artefacts of
order~$a$ in our data. We have addressed this issue by studying the
influence of different choices for improvement coefficients on our
results. To this end we have repeated the complete analysis using
non-perturbative values for $\ca$ and the combination $\ba-\bp$
obtained in the quenched approximation~\cite{impr:pap3,impr:bAbP} at a
similar value of the lattice spacing, $a\approx0.1\,\fm$. We found
that the resulting variation in the estimates for~$\alpha_5^{(2)}$ and
$(2\alpha_8^{(2)}-\alpha_5^{(2)})$ is typically a factor~10 smaller
than the statistical error. Thus we conclude that the influence of
improvement coefficients on our results is very weak indeed.

As a final comment we point out that we have not taken finite volume
effects into account in our error estimates, because different lattice
sizes were not considered in our study (unlike in earlier
simulations~\cite{dspect:ukqcd98}). However, since $L\mps=4.59$ at the
reference point and $L\mps=3.68$ at the lightest valence quark mass,
one may not be totally convinced that such effects may be entirely
neglected. We stress, though, the the definition of $\RFG$ and $\RF$
implies that only the {\it relative\/} finite-size effects between
hadronic quantities is relevant. Thus, as long as the mass
parameter~$x$ does not differ too much from unity, one can reasonably
expect that finite-volume effects largely cancel in the ratios used to
determine the low-energy constants. Finite-size effects in
pseudoscalar masses and decay constants have also been studied in ChPT
\cite{GaLe:87,GaLe:87b,chir:BerGol92,chir:Sha92}. These calculations
indicate that the typical relative finite-volume effect in $\mps$ and
$\Fp$ between our reference point and the smallest quark mass is less
than 1\%.

\section{Discussion and outlook \label{sec_disc}}

After combining the different systematic errors in quadrature we
obtain as our final results in two-flavour QCD:
\bea
   \alpha_5^{(2)} &=& 1.22\,^{+0.11}_{-0.16}\,\hbox{(stat)}\,
                          ^{+0.23}_{-0.26}\,\hbox{(syst)}
\label{eq_a5_res} \\
   (2\alpha_8^{(2)}-\alpha_5^{(2)}) &=& 0.36\pm0.10\,\hbox{(stat)}\,
                          \pm0.22\,\hbox{(syst)}
\label{eq_2a8a5_res} \\
   \alpha_8^{(2)} &=& 0.79\,^{+0.05}_{-0.07}\,\hbox{(stat)}\,
                          \pm0.21\,\hbox{(syst)}, \label{eq_a8_res}
\eea
where eqs.~(\ref{eq_a5_res}) and~(\ref{eq_2a8a5_res}) have been
combined to produce the result for $\alpha_8^{(2)}$.

We can now investigate the dependence of the low-energy constants on
the number of dynamical quark flavours. In the quenched
approximation~\cite{mbar:pap4} (i.e. for $\Nf=0$) it was found
that\footnote{When extracting the result for $\alpha_8^{(0)}$ it was
assumed that the coefficients multiplying quenched chiral logarithms
were set to $\delta=0.12, \alpha_\Phi=0.0$.}
\bea
   \alpha_5^{(0)} &=& 0.99\pm0.06\,\hbox{(stat)}
   \pm0.2\,\hbox{(syst)}  \label{eq_a5_qu} \\
   \alpha_8^{(0)} &=& 0.67\pm0.04\,\hbox{(stat)}
   \pm0.2\,\hbox{(syst)}. \label{eq_a8_qu}
\eea
A comparison with \eq{eq_a5_res} and~(\ref{eq_a8_res}) then shows that
$\alpha_5^{(2)}$ and $\alpha_8^{(2)}$ are larger than their quenched
counterparts by 23\% and 18\% respectively. We can thus conclude that
the $\Nf$-dependence of the low-energy constants is fairly weak:
variations between the quenched and two-flavour theories are about as
large as the error due to neglecting higher orders. 

Although there is {\it a priori\/} no reason why the weak
$\Nf$-dependence should extend to the physical three-flavour case, it
is still instructive to compare eqs.~(\ref{eq_a5_res})
and~(\ref{eq_a8_res}) with phenomenological values of the low-energy
constants. It then becomes obvious from eqs.~(\ref{eq_a5_phen})
and~(\ref{eq_a7a8_std}) that our results for $\alpha_5^{(2)}$ and
$\alpha_8^{(2)}$ are compatible with the standard estimates found in
the literature. By contrast, our numerical data for $\RF$ and $\RFG$
suggest that a large negative value for $\alpha_8$, which is required
for the scenario of $\mup=0$ (see \eq{eq_a7a8_mup0}), is practically
ruled out. Thus, provided that the quark mass behaviour in the
physical three-flavour case is not fundamentally different, the
possibility of a massless up-quark is strongly disfavoured.

\begin{figure}[tb]
\hspace{0cm}
\vspace{-8.5cm}

\centerline{
\leavevmode
\psfig{file=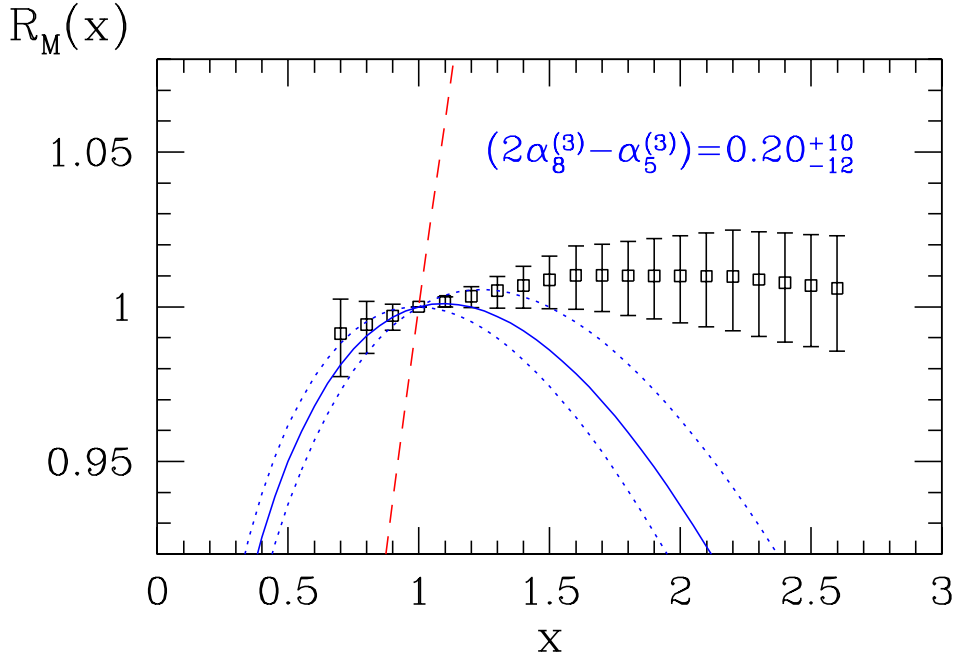,width=18cm}
}
\vspace{-0.3cm}
\renewcommand{\baselinestretch}{0.95}
\caption{\small Data for $\RFG$ (VV case) and the curve which results if
$\Nf=3$ in the determination of $(2\alpha_8^{(3)}-\alpha_5^{(3)})$
(solid line). Dotted lines represent the statistical uncertainty. The
dashed curve corresponds to $(2\alpha_8^{(3)}-\alpha_5^{(3)})=-2.3$, a
value which is consistent with the hypothesis of a massless
up-quark.\label{fig_RM_Nf3}}
\end{figure}

By how much does one expect the mass dependence of $\RFG$ and $\RF$ to
differ between $\Nf=2$ and~3\,? Ultimately this must be answered by a
direct simulation of the three-flavour case. For the time being we
have to be content with the following {\it gedanken}
simulation. Suppose that we had analysed our $\Nf=2$ data under the
erroneous assumption that they had been obtained in the physical
three-flavour case. We would then have set $\Nf=3$ in
eqs.~(\ref{eq_RFPS_VV}) and~(\ref{eq_RFPS_VS1}) to extract
$\alpha_5^{(3)}$, giving
$\alpha_5^{(3)}=0.98\err{11}{16}\err{23}{26}$. This value can be
inserted into the expression for $\FK/\Fpi$ in the physical
theory~\cite{chir:GaLe2}, to yield
\be
   \frac{\FK}{\Fpi} = 1.247\,^{+0.009}_{-0.013}\,\hbox{(stat)}
   \,^{+0.019}_{-0.021}\,\hbox{(syst)},
\ee
which is in fair agreement with the experimental result
$\FK/\Fpi=1.22\pm0.01$. This shows that the experimental value can
only be reproduced if the quark mass dependence of $\RF$ in the
physical case is not much different from that encountered in our
simulations for $\Nf=2$. In other words, it is reasonable to assume
that the mass dependence of $\RF$ is only weakly distorted by
neglecting the dynamical quark effects due to a third flavour.

Similarly we can apply the expressions for $\RFG$ for $\Nf=3$ to our
data, which gives
$(2\alpha_8^{(3)}-\alpha_5^{(3)})=0.20\err{10}{12}\err{22}{23}$. The
corresponding curve is shown in Fig.~\ref{fig_RM_Nf3}. The first-order
mass correction $\Delta_{\rm M}$ is then obtained as
\be
   \Delta_{\rm M} = -0.04\,^{+0.05}_{-0.06}\,\hbox{(stat)}
   \pm0.11\,\hbox{(syst)},
\ee
which is consistent with Leutwyler's estimate (see
\eq{eq_DeltaM_est}). We emphasise that this does not represent a
reliable result for $\Delta_{\rm M}$ derived from first principles.
Nevertheless, the above discussion shows that there are examples which
support the idea that the gross features of the mass dependence do not
differ substantially in the two- and three-flavour cases. On the basis
of this assumption one may conclude that the correction factor
$\Delta_{\rm M}$ is indeed small, ruling out the scenario of a
massless up-quark. As a further illustration we have included in
Fig.~\ref{fig_RM_Nf3} the curve which one would expect if $\mup=0$, by
taking the central values for $\alpha_5$ and $\alpha_8$ from
eqs.~(\ref{eq_a5_phen}) and~(\ref{eq_a7a8_mup0}).

The first priority for future work is undoubtedly the application of
the method to simulations employing $\Nf=3$ flavours of dynamical
quarks, and the extension of the quark mass range towards the chiral
regime. While efficient simulations of QCD with odd $\Nf$ and light
dynamical quarks represent an algorithmic challenge, some efforts in
this direction have already been made~\cite{dspect:MILC_3f}. It would
also be interesting to extend applications to the case of flavour
singlets, which allow a determination of $\alpha_7$~\cite{ShaSho_L7},
i.e. another low-energy constant afflicted with the KM
ambiguity. Methods to improve the notoriously bad signal/noise ratio
in flavour-singlet correlators have been
developed~\cite{sing:ukqcd_eta,sing:ukqcd_mix}, so that there are good
prospects for a successful implementation.

\section*{Acknowledgements}

We acknowledge the support of the Particle Physics \& Astronomy
Research Council under grants GR/L22744 and PPA/G/O/1998/00777. We are
grateful to the staff of the Edinburgh Parallel Computing Centre for
maintaining service on the Cray T3E.

\bibliography{biblist}        
\bibliographystyle{h-elsevier}   

\end{document}